\documentclass[conference]{IEEEtran}
\IEEEoverridecommandlockouts
\usepackage{cite}
\usepackage{amsmath,amssymb,amsfonts}
\usepackage{algorithmic}
\usepackage{graphicx}
\usepackage{textcomp}
\usepackage{xcolor}
\usepackage{hyperref}

\usepackage{booktabs}
\usepackage{adjustbox}
\usepackage{multirow}
\usepackage{hhline}
\usepackage{makecell}
\usepackage{tabularx}

\newcommand{\rom}[1]{\uppercase\expandafter{\romannumeral #1\relax}}
\def\BibTeX{{\rm B\kern-.05em{\sc i\kern-.025em b}\kern-.08em
    T\kern-.1667em\lower.7ex\hbox{E}\kern-.125emX}}
\begin{document}

\title{Multi-view user representation learning for user matching without personal information\\
}
\author{\IEEEauthorblockN{1\textsuperscript{st} Hongliu CAO}
\IEEEauthorblockA{\textit{Amadeus SAS}\\ \textit{hongliu.cao@amadeus.com}}

\and
\IEEEauthorblockN{2\textsuperscript{nd} Ilias El Baamrani}
\IEEEauthorblockA{\textit{Ecole des Mines de Nancy} }
\and
\IEEEauthorblockN{3\textsuperscript{rd} Eoin Thomas }
\IEEEauthorblockA{\textit{Amadeus SAS} }
}

\maketitle

\begin{abstract}
As the digitization of travel industry accelerates, analyzing and understanding travelers' behaviors becomes increasingly important. However, traveler data frequently exhibit high data sparsity  due to the relatively low frequency of user interactions with travel providers. Compounding this effect the multiplication of devices, accounts and platforms while browsing travel products online also leads to data dispersion. To deal with these challenges, probabilistic traveler matching can be used. Most existing solutions for user matching are not suitable for traveler matching as a traveler's browsing history is typically short and URLs in the travel industry are very heterogeneous with many tokens. To deal with these challenges, we propose the similarity based multi-view information fusion to learn a better user representation from URLs by treating the URLs as multi-view data. The experimental results show that the proposed multi-view user representation learning can take advantage of the complementary information from different views, highlight the key information in URLs and perform significantly better than other representation learning solutions for the user matching task.

\end{abstract}


\section{Introduction}
 
The travel industry is relying more and more on online solutions that make traveler’s experience more seamless and convenient. 
Traveler data are very rich during a traveler's journey from inspiration to post-travel. Search history prior to making a booking decision can be useful, for example, for a better understanding of traveler's behaviors during inspiration so that appropriate suggestions can be provided. Interactions with future bookings after a conversion also give great insights into how travelers interact with the products and prepare their trips.

	However, analyzing traveler’s data is very challenging due to the following reasons: 
	\begin{itemize}
	    \item Traveler’s data are sparse: Unlike other e-commerce solutions (e.g. Amazon, Alibaba, etc.) where many searches and purchases are made over a short period such as one month, a passenger may only reserve flight tickets once or twice a year \cite{cao2021destination}. 
	    \item Traveler's data are dispersed and fragmented: 
	    \begin{itemize} 
	    \item Across different platforms:  travelers often book flight tickets, trains, hotel rooms, or activities on different platforms/websites. 
	    \item Across different accounts: a user can have multiple accounts with different emails on different websites.
	    \item Across different devices or browsers: as the variety of devices increases, travelers nowadays often have multiple devices such as laptops, smartphones, tablets that can be used to search or book trips.
	    \end{itemize}
	\end{itemize}
Traveler's online behavior history is broken into many small non-connected pieces, which worsens the data sparseness problem. Therefore,  building an accurate traveler profile and understanding correctly a traveler's needs based on the dispersed data is difficult. 

The most direct way to deal with the data dispersion problem is to perform traveler matching, which builds a connection among traveler's different digital IDs. In the field of web content data mining, cross-device user matching has attracted a lot of attention as it is beneficial to both customers and the content providers \cite{inproceedings}.  The multitude of digital IDs per traveler mainly come from cookie cleaning and usage of multiple devices.   

In this work, we aim at matching travelers based on their search history from multiple online travel websites, tracked by anonymized cookie IDs. Recent data mining competitions have addressed similar problem using browsing logs of users' entire online activity. The state of the art solutions often use the sequence modeling over the user browsing logs, mostly in an unsupervised manner. However, these solutions are not well suited to travel industry data due to the sparsity and fragmentation present in the data. With the demise of 3rd party cookies, the data dispersion problem will likely appear in many other industries too.  

Unlike existing solutions that focus primarily on the final classification task of user matching, we focus instead on learning a good user representation. The main contributions of this study are:
\begin{itemize}
    \item a novel solution is proposed to address the sparsity and dispersion in the data using similarity based multi-view information fusion
    \item the proposed solution treats URLs as multi-view data, which can extract useful information from URLs efficiently and highlight the key information in URLs for the user matching task
    \item the proposed solution can take advantage of complementary information from different information sources as well as from different learning schemes to learn better user representations
\end{itemize}

The remainder of this paper is organized as follows: the background and related works are introduced in Section \rom{2}. In Section \rom{3}, the travel industry dataset used in this work as well as the overview of the proposed framework are introduced. We describe the multi-view user representation learning in Section \rom{4} and the inference model design in Section \rom{5}. The ablation study results are shown in Section \rom{6}. Finally, the conclusion and future works are given in Section \rom{7}.

\section{Related works} 
In the literature, the solutions for user matching can be divided into two main categories: deterministic user matching and probabilistic user matching. 
The straightforward approach to solve this problem is to rely on deterministic matching using robust identifiers like passport number, phone number, name, etc\cite{brookman2017cross}. However, deterministic matching is not always feasible. For example, during the inspiration phase, when travelers are searching for a destination/flight, the search history rarely contains Personal Identifiable Information (PII). Besides, user privacy is becoming a growing concern for customers \cite{lin2021cross}. In order to be aligned with the GDPR regulation and protect user privacy, deterministic matching using PII will likely become less prevalent. 

In this work, we focus on probabilistic matching, an alternative strategy to link users that does not require PII information. Probabilistic matching involves user representation learning, candidate selection and pairwise classification using physical and behavioral features \cite{inproceedings}. The former is more related to intrinsic device specificities such as device type, browser configuration, operating system, IP location, device time zone, among others. Although this type of information can provide straightforward clues about the user, they can easily become obsolete over time, especially due to frequent system updates \cite{inproceedings}. The behavioral features contain information about the preferences and behaviors that the user may engage in \cite{jiang2016little}. The most common behavioral features are the URLs of a user's browsing history with their respective timestamps. 

In the context of the annual International Conference on Information and Knowledge Management (CIKM), a competition under the name of cross device entity linking challenge was held to match users using their browsing history. The dataset used is published by the Data-Centric Alliance (DCA) \cite{Codalab} and contains anonymized user browsing logs. The leading solutions of this contest showed that unsupervised natural language models  have very good performances in terms of user representation learning. The winning team \cite{cikm1} and other challengers such as \cite{cikm2} used a Doc2vec model \cite{doc2vec} to embed sequences of URLs visited by users, considering the URL tokens as words in a sentence. However, such sequence modeling is not adapted to traveler data because of data sparsity (the average sequence length in travel data is around 6 times shorter than the DCA data). The runner up solution\cite{cikm3} used a TF-IDF weighting scheme and focused on feature engineering by introducing temporal correlations between browser logs, overlapping session activity, and analysis of key URLs whose presence gives strong matching probabilities. There are also various studies listed in the survey \cite{inproceedings} using NLP solutions along with different similarity measures to learn embeddings that represent user interests. 
A recent article \cite{cikm2021} also addressed the same topic based on the DCA dataset. In addition to Doc2vec, they used TF-IDF to represent browser logs but enhanced it with latent semantic indexing \cite{lsi}. This proposal improved the recall of the winning solution by 22\% despite a 4\% loss in the precision. 

Existing solutions often focus on unsupervised sequence embedding which are not necessarily relevant for the final user matching task. Generating embeddings with a supervised approach can provide classification information that can not be learned in the unsupervised setting. One way of learning the supervised user representation is through Siamese Networks \cite{Koch2015SiameseNN}, which have been used extensively in image similarity \cite{chen2022siamese, rafael2021re} and Nature Language Processing (NLP) tasks such as sentence embedding learning\cite{sbert, chicco2021siamese}. The researchers from \cite{siamesecikm} used Siamese networks to get richer embeddings than URL sequences, which improves the F1 score of the winning solution \cite{cikm1} by 4\%. 

In summary, most state of the art solutions proposed for the user matching task are based on the DCA dataset, while the travel industry data are different and more challenging due to the inherent data sparsity and data dispersion. The URLs in the travel industry are relatively long (in terms of number of tokens in the URL) and heterogeneous, however the sequence length of URLs in the travel industry is much shorter than in the DCA dataset, which means that the sequence models used in the previous solutions are less suitable for travel data.   
In the following sections, we propose multi-view information fusion based user representation learning using both supervised and unsupervised learning methods to efficiently extract useful information from URLs by treating URLs as multi-view data.

\section{Dataset}
The dataset used in this work contains browsing logs on different online websites during 9 months. The collected data contain 87 574 anonymized cookieIDs. Each traveler can have multiple cookieIDs due to cookie cleaning, multiple browsers, multiple devices or other reasons.  The anonymized emails are used to represent the ground truth data. These 87 574 anonymized cookieIDs in the dataset corresponds to 41 732 emails.  
Each cookieID is associated with a list of events that describe all the actions and interactions that the traveler has done. Each event is composed of physical and behaviour data including:

\textbf{Physical features:} User agent, Browser, Operation System (OS), Geoip city, Website language

\textbf{Behavioural features:} URL, Timestamp, Number of passengers, Product type, Product name,  Search level.

All the features are anonymized in this work. Table \ref{tab:dataset} illustrates the general statistics information about the dataset. It can be seen that the average number of unique URLs for each user (sequence length) in the travel industry data is 12.9, much shorter than 81 in the CIKM dataset.  

\begin{table}[htbp]
\caption{Dataset statistics}
\centering
\begin{tabularx}{250pt}{X l}
\toprule
Data                           & Number    \\ \midrule
\#Anonymized cookieIDs            & 87 574    \\
\#Anonymized emails                       & 41 732    \\
\#matching pairs               & 56 905    \\
\#Anonymized URLs                         & 1 131 679 \\
\#avg unique URLs per cookieID & 12.9     \\
\#Anonymized useragents                   & 23872     \\
\#Anonymized departure airports           & 1730      \\
\#Anonymized destination airports         & 2115      \\
\#Anonymized products types               & 4         \\
\#Anonymized geoip countries              & 197       \\
\#Anonymized geoip cities                 & 14880     \\ \bottomrule
\end{tabularx}%

\label{tab:dataset}
\end{table}

\begin{figure}
    \centering
    \includegraphics[width=0.49\textwidth]{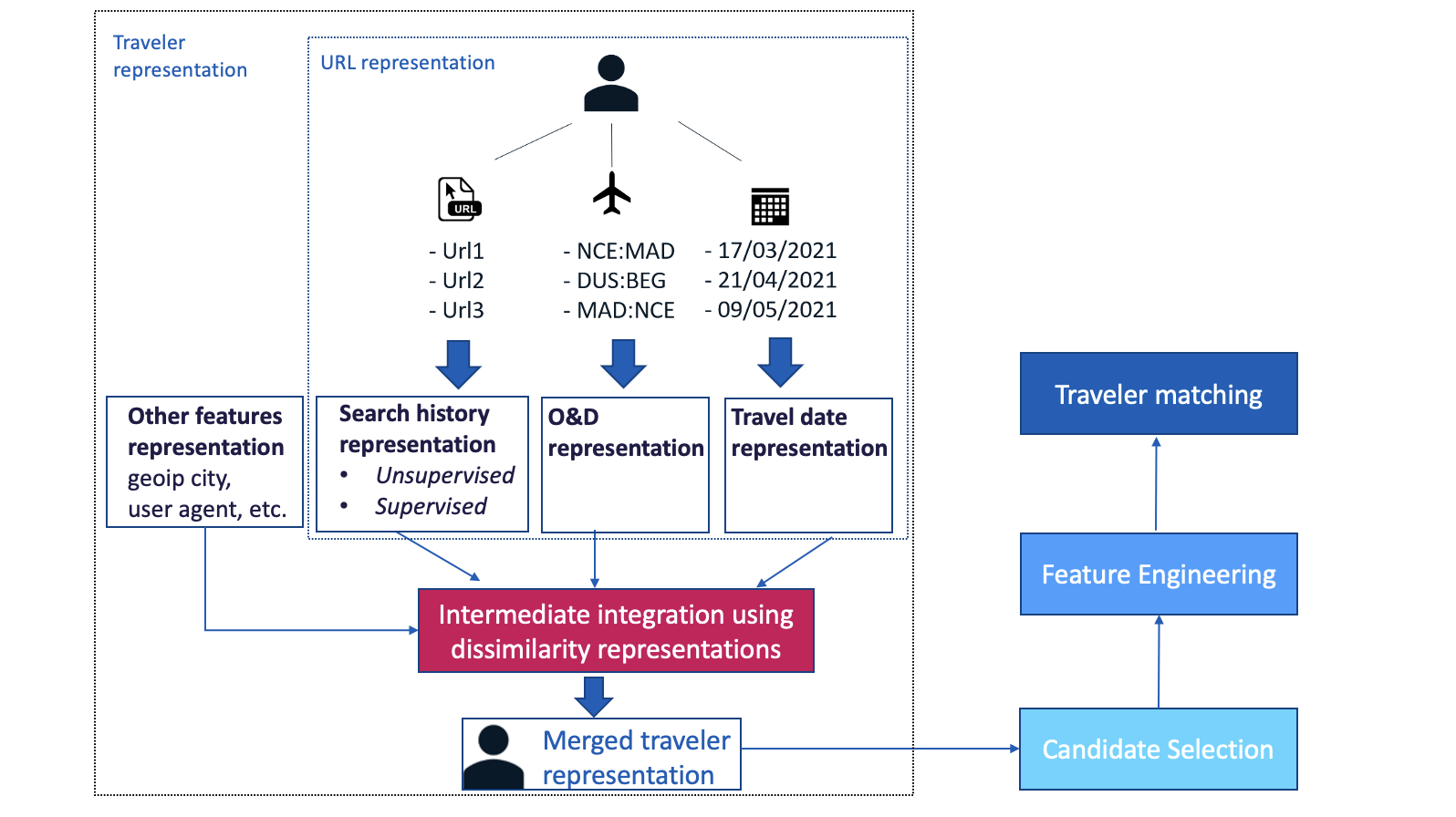}
    \caption{The overall workflow of the proposed solution. Step 1: Traveller representation learning by merging the representations learnt from URL and other features. Step 2: Candidate selection ( filter out irrelevant candidates). Step 3: Pairwise feature engineering. Step 4: Pairwise classification on selected candidate pairs. }
    \label{fig:workflow}
\end{figure}
The overall framework of the proposed solution is shown in Figure \ref{fig:workflow}. The first step is to learn a good user/traveler representation. Based on the user representation, we can select the most relevant candidates among the big candidate pool during Step 2. In step 3,  pairwise features are engineered for the final classification task. In step 4, the selected candidates are compared with the query user to identify if this candidate is a match or not. In the literature, most studies only evaluate the performance of the final pairwise classification task and ignore the evaluations of the user representation learning and candidate selection, which makes their insights and conclusions less interpretive. In this work, a step by step explanations and evaluations are introduced in the following sections. 

For all the experiments in this work, we randomly select 70\% emails to form training set and 30\% to form test set with 5 different repetitions and report the average performance as well as the standard deviation. 

\section{User representation learning}
In this section we seek to build travelers' representation that incorporates their online behaviours and preferences for the task of user matching. To make sure that the learnt user representation is useful for the candidate selection and pairwise classification task, the TopN accuracy is used to in this work to evaluate the quality of the user representations. Given each user vector representation $u$ of dimension $d$, the cosine similarity measure is used to build a user$\times$user similarity matrix $S = [s_{u,v}]^{n\times n}$ of $n$ users where $s_{u,v}$  is the similarity value between the vector representation of user $u$ and user $v$. 
Based on the user$\times$user similarity matrix, we evaluate the quality of the user representation  by calculating, on average, the percentage of correct matches identified in the TopN most similar users (TopN accuracy). For example, if the top 5 most similar users of the user $u$ from the similarity matrix are users [a,e,p,l,h] and the ground truth are: [e,z],the Top5 accuracy is 0.5 as we identified only half of the ground truth in the top 5 most similar users.

\subsection{User representation learning from URL}
URLs browsed by users are the most important information used in the literature. The URLs in travel industry can be very long as they usually contain a lot of information such as origin, destination, departure date, return date, product type, product ID, search level, passenger number, booking class, sorting type and so on. They are also very heterogeneous because they don't have necessarily the same structure on different websites or at different search levels (sales funnels). 

As mentioned above, the main challenge in the travel industry is the data sparsity problem: the average URL sequence length for each user is around 13 compared to 81 in the CIKM dataset. The shortest sequence length in our dataset is 2. To deal with this limitation, we put all the tokens
extracted from the URLs browsed by each cookieID together to form the pool of tokens instead of using sequences of different levels of URL as in \cite{cikm1}.  

\subsubsection{Unsupervised user representation learning}
Unsupervised learning is the most widely used in the literature to extract embedding from URLs \cite{inproceedings}. Following the suggestions of the previous works, Term Frequency 
Inverse Document Frequency (TF-IDF) is used to learn the user representation from URLs. For our experiments, we keep the feature dimension at 1000 because there is no significant difference in terms of TopN accuracy between 1000 feature dimension and higher feature dimensions. 

\subsubsection{Supervised user representation learning}

\begin{figure}[htbp]
\centering
\includegraphics[width=0.25\textwidth]{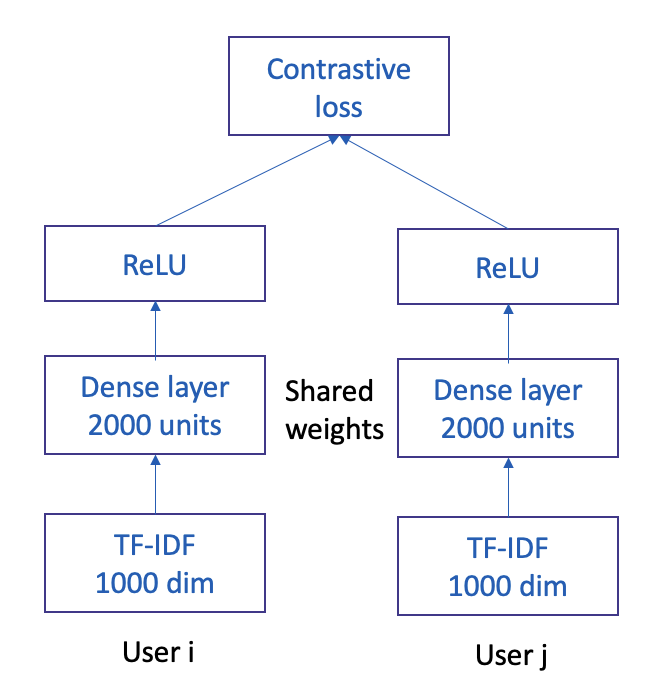}
\caption{Siamese network architecture}
\label{fig:d}
\end{figure}

The supervised representation learning can take advantage of the ground truth matching pairs information to learn more relevant representations for the user matching task. 
In this work, Siamese Network is used to learn the supervised embedding. 
The proposed Siamese network consists of one dense layer followed with ReLU activation as shown in Figure \ref{fig:d}. The input of the Siamese Network is the TF-IDF vector obtained above. 
The loss function used in this work is the contrastive loss \cite{Hadsell2006DimensionalityRB}, which minimizes intra-class embeddings distance (matched users) and maximizes inter-class distance (non matched users). The contrastive loss is defined as Equation \ref{contrastive}, where the margin $m$ prevents dissimilar examples from getting too close.

\begin{equation}
Loss=
\left\lbrace
\begin{array}{ccc}
\frac{1}{2} \lVert f_i-f_j \rVert_{2}^2 & \text{if} & y_{ij} = 0 \\ \\

\frac{1}{2}\max(0,m- \lVert f_i-f_j \rVert_{2}^2) & \text{if} &  y_{ij} = 1
\end{array}\right.
\label{contrastive}
\end{equation}

where $f_i$ and $f_j$ are the learnt URLs embedded representations for user $i$ and $j$ respectively. The $y_{ij}$ represents the training label: it is set to 1 if the pair of users don't match and 0 otherwise.

Figure \ref{fig:d} illustrates the Siamese network architecture used in this work. After trying different layer numbers, we find out that one dense layer is sufficient to learn a good URL representation from the TF-IDF input. The margin used in the contrastive loss function is set to 3 (through cross validation on training data) which gives the best results in our experiments. ADAM \cite{Kingma2015AdamAM} is used as the optimizer with a learning rate of 0.001. 
During training, the Siamese Network is fed with challenging pairs so that it can be  efficient in identifying same users and distinguishing different (but similar) users. To form the training pairs, the real matches are selected as positive pairs. In terms of negative pairs sampling, we select the Top K most similar samples based on the cosine similarity values on the unsupervised TF-IDF representation to form the negative pairs. In this way, the selected negative pairs are challenging to distinguish.   After testing different values between 3 and 50, we fix K as 40 using cross-validation over training data.

To compare the performances between unsupervised  and supervised representation learning, we randomly select 70\% of the emails in the dataset for training and the remaining 30\% for testing over different repetitions and report the TopN accuracy (based on the cosine similarity values calculated between user representations). We also compare the performances with the best solution from CIKM \cite{cikm1}, where they used TF-IDF for each level in the URL hierarchy (i.e., a/b/c becomes [a, ab, abc]) and multi-level Doc2Vec. The experimental results are shown in Table \ref{tab:URLs}. Compared to the multi-level TF-IDF, our unsupervised representation (TF-IDF with simple URL tokens pool) has much better performance. This confirms our observation that the URLs in the travel industry are longer and more heterogeneous (the same level in the URL does not always contains the same information). It can also be told from Table \ref{tab:URLs} that the Doc2Vec used in \cite{cikm1} has the worst performance, which shows that treating each level in the URL separately with Doc2Vec is not suitable for the travel data due to the data sparsity and short sequence length. Finally, compared to the unsupervised representation learning, the supervised representation has significantly better performance.  

\begin{table}[]
\centering
\caption{TopN accuracy of the URL representations on test set. }
\label{tab:URLs}
\resizebox{3.6in}{!}{%
\begin{tabular}{|l||l||l||l||l|}
\hline
Accuracy & Multi-level TF-IDF  \cite{cikm1} & Doc2Vec \cite{cikm1}  & Unsupervised    & Supervised      \\ \hline
Top3           & 22.38\%$\pm$0.14 & 8.96\%$\pm$0.15  & 37.37\%$\pm$0.16 & 50.47\%$\pm$0.29 \\ \hline
Top5           & 24.52\%$\pm$0.18 & 10.79\%$\pm$0.12 & 41.51\%$\pm$0.25 & 54.66\%$\pm$0.21 \\ \hline
Top10          & 27.82\%$\pm$0.21 & 13.79\%$\pm$0.13 & 47.22\%$\pm$0.22 & 60.05\%$\pm$0.17 \\ \hline
Top18          & 30.96\%$\pm$0.18 & 16.77\%$\pm$0.12 & 52.27\%$\pm$0.22 & 64.32\%$\pm$0.15 \\ \hline
Top50          & 37.75\%$\pm$0.18 & 23.33\%$\pm$0.19 & 61.50\%$\pm$0.23 & 71.29\%$\pm$0.17 \\ \hline
\end{tabular}
}
\end{table}

\subsubsection{Representation learning from highlights in the URL}
As we mentioned above, the URLs in travel industry are very long and contain a lot of information, among which the most important ones are the Origin and Destination (OnD) and travel dates as they reflect the traveler's preferences.
The question we raise here is: does the previously learnt representation from the entire URLs contain and highlight all these important information? To answer this question, the anonymized Origin and Destination (OnD) and the anonymized travel dates are extracted from the URL and two more user representations are learnt:

\textbf{OnD preference representation:} During several browsing sessions, each traveler usually looks for several  destinations. For each cookieID, we construct the document containing all searched origins and destinations and use TF-IDF to learn the OnD representation. 

\textbf{Dates preference representation:} This concerns the departure and return dates of the searched trip (in the case of a round trip). A traveler looking for a flight for a given date on her/his phone might look for the same dates on her/his computer. Therefore we put all the searched dates in the same document for each cookieID and learn the Dates preference representation with TF-IDF. 

The TopN accuracy of OnD preference representation and Dates preference representation are shown in Table \ref{tab:OnDs}. From the results, it can be seen that the OnD preference representation has better performance than the Dates representation, but still worse that the representation learnt from the entire URL. 

\begin{table}[htbp]
\caption{TopN accuracy of other feature representations on test set. The mean and std are shown for the four features}
\label{tab:OnDs}
\resizebox{3.5in}{!}{%
\begin{tabular}{|l||l||l||l||l|}
\hline
Accuracy & OnD             & Dates    & User agent      & Geoip city      \\ \hline
Top3           & 32.01\%$\pm$0.12 & 10.20\%$\pm$0.13 & 3.78\%$\pm$0.09 & 21.88\%$\pm$0.14 \\ \hline
Top5           & 38.32\%$\pm$0.25 & 12.48\%$\pm$0.14 & 4.85\%$\pm$0.11 & 26.31\%$\pm$0.14 \\ \hline
Top10          & 47.26\%$\pm$0.23 & 15.87\%$\pm$0.16 & 6.41\%$\pm$0.15 & 32.53\%$\pm$0.25 \\ \hline
Top18          & 54.55\%$\pm$0.29 & 19.18\%$\pm$0.09 & 8.04\%$\pm$0.21 & 37.57\%$\pm$0.28 \\ \hline
Top50          & 66.53\%$\pm$0.29 & 25.88\%$\pm$0.15 & 11.66\%$\pm$0.29 & 45.47\%$\pm$0.37 \\ \hline
\end{tabular}
}
\end{table}

\begin{table*}[htbp]

\centering
\caption{TopN accuracy for the merged representations. The mean and std are shown for the differents fusions involving five features}
\label{tab:fusion}
\begin{adjustbox}{width=1\textwidth} 
\begin{tabular}{|l||l||l||l||l||l|}
\hline
Model/TopN accuracy                                          & Top3            & Top5            & Top10           & Top18           & Top50          \\ \hline

Supervised   URL                                                             &
50.47\%$\pm$0.29 & 54.66\%$\pm$0.21 & 60.05\%$\pm$0.17 & 64.32\%$\pm$0.15 & 71.29\%$\pm$0.17
\\ \hline
Supervised   URL+OnD                                                            & 57.20\%$\pm$0.14 & 62.10\%$\pm$0.26 & 68.25\%$\pm$0.17 & 72.90\%$\pm$0.19 & 79.97\%$\pm$0.31 \\ \hline
Supervised   URL+OnD+Dates                                                      & 60.75\%$\pm$0.17 & 64.86\%$\pm$0.12 & 69.91\%$\pm$0.09 & 73.76\%$\pm$0.10 & 79.51\%$\pm$0.21 \\ \hline
Supervised   URL+OnD+Dates+Geoipcity                                            & 67.03\%$\pm$0.24 & 70.65\%$\pm$0.27 &74.99\%$\pm$0.27 & 78.36\%$\pm$0.18 & 83.46\%$\pm$0.32 \\ \hline
Supervised   URL+OnD+Dates+Geoipcity+User agent                                  & 67.48\%$\pm$0.25 & 71.13\%$\pm$0.24 & 75.51\%$\pm$0.28 & 78.80\%$\pm$0.22 & 84.02\%$\pm$0.14 \\ \hline
\textbf{Supervised   URL+OnD+Dates+Geoipcity+User agent+Unsupervised URL}                       & \textbf{69.99\%$\pm$0.28} & \textbf{73.71\%$\pm$0.26} & \textbf{78.14\%$\pm$0.23} & \textbf{81.45\%$\pm$0.25} & \textbf{86.49\%$\pm$0.16} \\ \hline
\end{tabular}
\end{adjustbox}
\end{table*}

\subsection{User representation learning from other features}
Apart from the representations learnt from URLs, we also learn the user representation from user agent and geoip city as they provide complementary information to URLs. 

\textbf{User agent} provides information on the configuration used by the user. An example of the user agent information can be: "Mozilla/5.0 (Linux; Android 10; LIFETAB E1080X) AppleWebKit/537.36 (KHTML, like Gecko) Chrome/89.0.4389.105 Safari/537.36". The user agents can be helpful in identifying users who clean their cookies as they usually still use the same device and thus have more or less similar user agents. 
We anonymize and tokenize the users agents and construct the "document" of all user agents for each user. TF-IDF is then used to learn the user agent representation. 

\textbf{Geoip city} indicates the city of the user's IP location when searching on the website. 
This feature can be very informative for the user matching task as users using multiple devices or cleaning their cookies are more likely to stay in the same town or city. 
We build a document of the anonymized cities where the user is located during all her/his online browsing history and build the Geoip city representation with TF-IDF. 

The TopN accuracy of user agent preference representation and Geoip city preference representation are shown in Table \ref{tab:OnDs}. It can be told that geoip city representation is much more informative than user agent representation. The main reason is that more people share the same user agent than 
 sharing the same geoip city. 

\subsection{User representation fusion}
In the works of \cite{cao2018improve, cao2019random, cao2021novel}, the authors find out that the dissimilarity based intermediate integration can efficiently take advantage of the complementary information among different views. In this work, each user representation learnt in the previous section can be seen as a view. We build a similarity matrix using the cosine similarity on each user representation including URL, OnD, Dates, User agent and Geoip city representations. To merge these heterogeneous representations, the average of these similarity matrices are calculated. 

The TopN accuracy of the representation fusion is shown in Table \ref{tab:fusion}. We gradually add each representation on top of the supervised representation of URLs (the best performing individual representation). When the OnD preference representation is added to the Supervised representation, the TopN accuracies are increased by around 7\% for all Ns. When the Dates preference representation is added, the performance is further improved. The results indicate interesting findings: even though the OnD and Dates information are already included in the URLs, highlighting their representation still adds complementary information to the supervised URL representation. This result provides the answer to the question we raised in the previous section (Does the user representation learnt from the entire URLs contain and highlight all these important information such as OnD and Dates?).  When adding Geoip city representation which is not contained in the URLs, the Top3 and Top5 accuracies are further improved by around 6\%. However, adding User agent information only improves slightly the quality of the merged user representation. Finally, we also add the unsupervised representation of the URLs, the TopN accuracies are improved again by 2-3\%. 

From the experiment results, we can get the following conclusions. 1. In terms of representation learning from URLs, the supervised learning has better performance than the unsupervised learning, but the unsupervised representation still contains useful complementary information. 2. Highlighting the important information in the URLs such as OnD and Dates can also help to improve the merged user representation quality. 3. Adding extra information such as User agent and Geoip city into the merged URL representation using the similarity based information fusion is helpful.   

\subsection{Candidate selection}
A brute force approach to generate candidates pairs is to create all possible pair combinations between cookieIDs and then decide whether they constitute a match or not. This approach results in a very large number of candidate pairs when the number of users increases. In candidate selection, we introduce some filters to reduce as much as possible the size of the candidate pairs while preserving those that have the most chance to result in a match. 

\begin{table*}[t]
\centering
\caption{Precision, Recall, F1 score of the inference model on the test set. The mean and std are given for 5 random splits. A comparison is given between 6 different models used in the candidate selection.  }
\label{tab:pairwise}
\begin{adjustbox}{width=0.75\textwidth} 
\begin{tabular}{|l||l||l||l|}
\hline
Model/Metrics                                         & \multicolumn{1}{c||}{Precision} & \multicolumn{1}{c||}{Recall} & \multicolumn{1}{c|}{F1} \\ \hline
Supervised URL                                           & 81.21\%$\pm$1.88                        & 44.82\%$\pm$1.00                     & 57.74\%$\pm$0.82             \\ \hline
Supervised   URL+OnD                                        & 83.87\%$\pm$0.92                        & 50.15\%$\pm$0.97                     & 62.76\%$\pm$0.80             \\ \hline
Supervised   URL+OnD+Dates                                  & 83.52\%$\pm$0.67                        & 50.87\%$\pm$1.02                     & 63.22\%$\pm$0.82             \\ \hline
Supervised   URL+OnD+Dates+Geoipcity                        & \textbf{84.53\%$\pm$1.09}                        & 54.07\%$\pm$0.74                     & 65.96\%$\pm$0.74             \\ \hline
Supervised   URL+OnD+Dates+Geoipcity +User agent              & 84.16\%$\pm$0.37                        & 54.21\%$\pm$1.11                     & 65.94\%$\pm$0.86             \\ \hline
\textbf{Supervised URL +OnD+Dates+Geoipcity +User agent+Unsupervised URL} & 82.59\%$\pm$0.56                        & \textbf{56.05\%$\pm$0.92}                     & \textbf{66.78\%$\pm$0.80}             \\ \hline
\end{tabular}
\end{adjustbox}
\end{table*}
For each user, we take advantages of  the merged user$\times$user similarity matrix generated in the information fusion part (the aggregation of representations from URLs, OnD,  Dates, Geoip city and User agent) in the previous section, to select the top k most similar users as candidates.   We use the same number of candidates k during both training and test stages to have the same imbalance ratio for the pairwise classification task. Choosing a larger k  increases the size of the candidate pairs for each user (which also increases the computational cost) and increases the imbalance ratio (which makes the pairwise classification task harder), while choosing a smaller k decreases the chance of the correct pair being selected.  From the results in Table \ref{tab:fusion}, we can see that if we select 3 candidates for each user, the chance of the correct candidates being selected is 69.99\% (the Top3 accuracy) while if we select 50 candidates for each user, the chance of the correct candidates being selected is 86.49\% (the Top50 accuracy). In this work, we set $k=5$ based on the cross-validation performance on the training data. 

\section{Inference model}
In this work, we tackle the problem of traveler matching as a pairwise classification problem. Given a pair of cookieIDs, the objective is to determine whether they belong to the same user or not. In the previous section, we select k candidates for each cookieID. The next step is to build features for the pairwise classification task and choose the best classifier. 

\subsection{Pairwise feature engineering}
In this work, we design pairwise features tailored to the travelers' data. We categorize the features into 3 types: Traveler preference features, Traveler behavior features and Traveler similarity features. 

\subsubsection{Traveler preference features}

We assume that two cookieIDs that share similar preferences are more likely to belong to the same user. We therefore generate the following features:



\textbf{SamePassengers}: While searching for a flight, users usually indicate the number of passengers. We keep for each user the number of passengers queried the most and build a binary feature to tell if two users searched the same number of passengers. 

\textbf{JaccardPassengers}: the Jaccard similarity between the  number of passengers queried by two users.

\textbf{SameOS}: Binary feature that tells if a pair of cookieIds have the same OS.

\textbf{SameCity}: Considering the city where the user has done the most searches, we build a binary feature that tells if two cookieIDs are from the same city or not.

\subsubsection{Traveler behaviour features} 
We build the following pairwise features based on travel behaviours : 





\textbf{SearchLevel}: The search level indicates the hierarchical level of the sales funnel. To construct the pairwise feature, we average all the search levels for the cookieIDs in each pair and compute their absolute difference.

\textbf{TimeRange}:  Given the activity period (cookie lifetime) of the user $u$ and the activity period of the user $v$, we calculate the number of overlapped days as TimeRange feature.

\textbf{TimeToFlight}: We assume that users may have patterns regarding the search date with respect to the travel date. We thus calculate for each user the  differences between travel dates and search dates and compute their average. The absolute difference is then calculated for each pair of cookieIDs. 

\subsubsection{Traveler similarity feature}  
In terms of OnD, Dates, Geoip City and User agent, we calculate the Jaccard similarity between two users as the similarity features. As we already built the vector representations for these features in the previous section, we also calculate the cosine similarity over their vector representations between two users as pairwise features too. On top of these features, the following ones are also used:

\textbf{CosineURLs}: the Cosine similarity between the URL representations of user pairs. 

\textbf{Overall similarity}: In the previous section, the user similarity matrix based on the merged user representation is used for the candidate selection. Here, the pairwise similarity value is also used as a feature. 

\subsection{Pairwise classification}

After designing the pairwise features for the selected candidate pairs, we use H2O AutoML \cite{H2OAutoML20} framework to select the best ML model that optimizes the Area Under the Precision Recall Curve (AUCPR). The best model selected is the Gradient Boosted Machine (GBM), which has been very successful in  solving various classification tasks. We evaluate the precision, recall and F1 score on the test set and report the results in Table \ref{tab:pairwise}. We show the performances for the different levels of information fusion used in the user representation learning and candidate selection in comparison to the results in Table \ref{tab:fusion}.

\begin{table*}[htbp]
\centering
\caption{TopN accuracy of the reduced model. The mean and std are shown for  different fusions involving three features}
\label{tab:generalization-topN}
\begin{tabular}{|l||l||l||l||l||l|}
\hline
Model/TopNaccuracy                & \multicolumn{1}{c||}{Top3} & \multicolumn{1}{c||}{Top5} & \multicolumn{1}{c||}{Top10} & \multicolumn{1}{c||}{Top18} & \multicolumn{1}{c|}{Top50} \\ \hline
WebsiteDomain                    & 0.89\%$\pm$0.06               & 1.33\%$\pm$0.06               & 2.27\%$\pm$0.05                & 3.46\%$\pm$0.04                & 7.15\%$\pm$0.04                \\ \hline
OnD                               & 32.01\%$\pm$0.12                         & 38.32\%$\pm$0.25                         & 47.26\%$\pm$0.23                          & 54.55\%$\pm$0.29                          & 66.53\%$\pm$0.29                          \\ \hline
Dates                      & 10.20\%$\pm$0.13                         & 12.48\%$\pm$0.14                         & 15.87\%$\pm$0.16                          & 19.18\%$\pm$0.09                          & 25.88\%$\pm$0.15                          \\ \hline
OnD + Dates                & 50.21\%$\pm$0.23               & 54.62\%$\pm$0.23               & 60.22\%$\pm$0.22                & 64.69\%$\pm$0.29                & 71.80\%$\pm$0.21                \\ \hline
\textbf{OnD+Dates+WebsiteDomain} & \textbf{51.35\%$\pm$0.15}               & \textbf{55.73\%$\pm$0.23}               & \textbf{61.15\%$\pm$0.34}                & \textbf{65.40\%$\pm$0.34}                & \textbf{72.42\%$\pm$0.38}                \\ \hline
\end{tabular}
\end{table*}

Generally speaking, by comparing the results shown in Table \ref{tab:pairwise} and Table \ref{tab:fusion}, we can tell that the better quality the user representation has, the better pairwise classification performance the user matching task achieves. This is because better user representation results in better candidate selection and better similarity features used in the classification task. It's interesting to note that, the better user representation has no significant influence on the precision scores while the recall and F1 scores are significantly improved with better user representation. That is because recall is defined as True Positive / (True positive + False Negative) which represents how many correct matching pairs among the ground truth matching pairs can be identified. A better user representation means better candidate selection, where more matching pairs can be selected as candidates and more matching pairs can be identified. The experiment results confirm that learning a good user representation through multi-view information fusion is very helpful for the user matching task.

\section{Ablation study}

In this section, more experiments and analysis are done to study:
\begin{itemize}
    \item How the proposed solution would behave when less information is provided as in other travel industry datasets. 
    \item How the proposed solution would behave in the real world application scenario where the data are split by time period instead of  by ground truth user pairs. 
\end{itemize}

\subsection{Model utility study: generalization to reduced features}
Firstly, we propose to study to which extent our solution can be used and generalized to other  datasets in the travel industry with fewer features.  We summarize that most datasets in the travel industry would have the following minimum feature set including 
OnDs, Travel Dates, Website Domains (office IDs), Number of passengers and Language. 
The OnDs, Travel Dates and Website domains are used to learn the traveler representation. Only unsupervised representations are considered here. Siamese Network is not used on URLs domains as they contain only 58 tokens. The topN accuracy of the different levels of information fusion are given in table \ref{tab:generalization-topN}. There is a 19\% decrease of the topN accuracy with regards to the complete model which includes in addition  URLs, Geoip cities and User agents. Similar to the conclusions from the previous section, it can be told that the multi-view information fusion can improve the user representation quality significantly (e.g. the merged representation  increases the Top5 accuracy of the OnD representation by 17.41\%).

In the inference model, we use the following minimized pairwise features:
\{\textit{CosineDomain, CosineOnds, CosineDates, SameLanguage, JaccardOnd, SameLanguage, JaccardOnd, JaccardPassengers, TimetoFlight, Overall Similarity}\}. 
  Table \ref{tab:generalization-F1score} shows the results of the pairwise classification. Without surprise, We observe a decrease on the precision, recall and F1 scores. But we still 
 obtain the F1 score of 48.98\%. The experimental results indicate that: the proposed system still works well even with minimum features and simplified solution (without supervised representations).

\begin{table}[]
\centering
\caption{Precision, Recall and F1 score of the reduced model}
\label{tab:generalization-F1score}
\begin{tabular}{|l||l||l||l|}
\hline
Metrics & Precision   & Recall      & F1 score    \\ \hline
Performance        & 73.51\%$\pm$0.55 & 36.73\%$\pm$0.68 & 48.98\%$\pm$0.56 \\ \hline
\end{tabular}
\end{table}

\subsection{Time split}

In the experimental protocol used in the previous section and the literature, the data train-test split is based on the groudtruth matching user pairs.  However, a user's online activity is in reality collected chronologically, which means that a new cookie ID that has just been tracked has a potential match in the cookies collected before. 
Hence, we propose to evaluate the performance of the proposed model in a more realistic industrial application setting by splitting the data by time. We sort the cookieIDs chronologically and divide them into 70\% training data 30\% test data. The groudtruth matching user pairs  distribution is shown below:
\begin{itemize}
    \item \textbf{57\%} of  ground truth pairs are exclusively present in training data. 
    \item \textbf{16\%}  of ground truth pairs are exclusively present in test data.
    \item \textbf{27\%} of ground truth pairs are present in both training and test data.
\end{itemize}

From the above statistics, it can be seen that the user matching task is more difficult in the new data split protocol. During test time, as the groudtruth can be present in both training and test data, the search space becomes bigger. The best pairwise classification model  in the previous section is trained and tested again in the new protocol. 

Table \ref{tab:topNaccuracy-timesplit} shows the performance of the merged user representation. Compared to the results in Table \ref{tab:fusion},  we observe a 10\% drop in the topN accuracy on average due to this new splitting configuration, which is expected as we compare within users in the entire dataset instead of on the test dataset only. We follow the same protocol as in the previous section for the candidate selection, generating 5 candidates for each user and then generating the pairwise features for the classification task. The precision, recall and F1 scores are shown Table \ref{tab:timesplit-F1score}: the F1 score drops to 49.37\% under the new experiment protocol, which shows that the  more realistic new train-test protocol (splitting data by time) is much more challenging. 
\begin{table}[htbp]
\centering
\caption{TopN Accuracy of the Time split model. "All" includes all the 6 features in the information fusion }
\label{tab:topNaccuracy-timesplit}
\resizebox{3.5in}{!}{%
\begin{tabular}{|l||l||l||l||l||l|}
\hline
Model/Accuracy                                        & \multicolumn{1}{c||}{Top3} & \multicolumn{1}{c||}{Top5} & \multicolumn{1}{c||}{Top10} & \multicolumn{1}{c||}{Top18} & \multicolumn{1}{c|}{Top50} \\ \hline
\makecell{Supervised URL \\ +OnD+Dates+Geoipcity \\ +User agent+Unsupervised URL}  & 59.66\%                   & 63.25\%                   & 67.64\%                    & 71.07\%                    & 76.57\%                    \\ \hline
\end{tabular}
}
\end{table}

\begin{table}[htbp]
\centering
\caption{Precision, Recall and F1 score of Time split model}
\label{tab:timesplit-F1score}
\begin{tabular}{|l||l||l||l|}
\hline
Features/Metrics & Precision   & Recall      & F1 score    \\ \hline
All          & 66.76\% & 39.17\% & 49.37\% \\ \hline
\end{tabular}
\end{table}

\section{Conclusion and future works}

In this paper, we analyzed and summarized the specific data challenges faced in the travel industry, such as data sparsity and data dispersion, which makes many state of the art solutions less suitable for travel data. We use traveler matching application as an example to show how travel data can be very different from the data in other industries. For example, the sequence length in travel data is much shorter and the URLs in travel industry is longer and more heterogeneous. 

To deal with the user matching problem, we process the tokens in the URLs differently from the existing solutions to mitigate the data sparsity challenge and propose to use similarity based multi-view information fusion to learn a better user representation from anonymized URLs. The information fusion for URL representation is multi-level: 1. we learn both a supervised user representation and an unsupervised user representation from the entire URLs of the user's browsing history; 2 we extract specific tokens from URLs such as anonymized OnD and anonymized travel dates and learn the user representation for each of them to further highlight the key information in the URLs. From the experimental results, we can tell that merging the supervised and unsupervised URLs representations along with the URL highlight representations (OnD and travel Dates representation) can improve the user representation quality significantly. This result shows that one single representation learning method may not be able to extract all useful information from the long URLs in travel industry. To further improve the user presentation, we also merge Geoip city and User agent that are not presented in the URLs to insert more  complementary information into the final user representation. A better user representation results in better candidate selection, which improves the pairwise classification performance for user matching. Our experimental results show that a user representation with higher TopN accuracy can improve significantly the Recall and F1 scores of the pairwise classification task without hurting the precision score.  
Other than user matching application, the  user representations learnt in this work can also be used for user clustering, segmenting or profiling. It can also be used for recommender systems where the user$\times$user similarity matrix is needed.

However, there is still a lot of room for improvements. Firstly,  more effort could be done to improve the supervised representation learning such as designing a better Siamese network with neural architecture search and designing a better negative sampling strategy. 
In terms of the multi-view information fusion, we average the different $user\times user$ similarity matrices to obtain the merged representation in this work. We could improve the information fusion using weighted average instead by choosing proper weights for each representation as in \cite{cao2018dynamic}. Thirdly, we use AutoML in this work for the pairwise classification task as the main focus of this paper is to learn a better user representation. More studies can be done to design better pairwise features and better classifiers.

\bibliographystyle{ieeetr}
\bibliography{sample}

\end{document}